\documentclass[twocolumn,showpacs,preprintnumbers,amsmath,amssymb]{revtex4}
\usepackage{graphicx}% Include figure files
\usepackage{dcolumn}% Align table columns on decimal point
\usepackage{bm}% bold math

\usepackage{epsfig} \usepackage{changebar} \usepackage{amsmath}

\newcommand{\be}{\begin{equation}}
\newcommand{\ee}{\end{equation}}
\newcommand{\bea}{\begin{eqnarray}}
\newcommand{\eea}{\end{eqnarray}}

\newcommand{\avg}[1]{\langle\langle{#1}\rangle\rangle}

\begin{document}

\bibliographystyle{plain}

\preprint{}

\title{Critical threshold and dynamics of a general rumor model on
  complex networks}

\author{Maziar Nekovee}
\affiliation{
Complexity Research Group, BT Research,
Polaris 134, Adastral Park, Martlesham, Suffolk IP5 3RE, United Kingdom}

\author{Yamir Moreno}

\affiliation{Institute for Biocomputation and Physics of Complex Systems and
  Department of Theoretical Physics, University of Zaragoza, Zaragoza 50009,
  Spain}

\author{Ginestra Bianconi and Matteo Marsili}
\affiliation{The Abdus Salam International Centre for Theoretical
  Physics, Strada Costiera 11, 34014 Trieste, Italy} 

\date{\today}% It is always \today, today,

\begin{abstract}
  We derive mean-field equations that describe the dynamics of a general model
  of rumor spreading on complex networks, and use analytical and numerical
  solutions of these equations to examine the threshold behavior and dynamics
  of the model on random graphs, uncorrelated scale-free networks and
  scale-free networks with assortative degree correlations. We show that in
  both homogeneous networks and random graphs the model exhibits a critical
  threshold in the rumor spreading rate below which a rumor cannot propagate in
  the system. In the case of scale-free networks, on the other hand, this
  threshold becomes vanishingly small in the limit of infinite system size.  We
  find that the initial rate at which a rumor spreads is much higher in
  scale-free networks than in random graphs, and that the rate at which the
  spreading proceeds on scale-free networks is further increased when
  assortative degree correlations are introduced.  The impact of degree
  correlations on the final fraction of nodes that ever hears a rumor, however,
  depends on the interplay between network topology and the rumor spreading
  rate.  Our results show that scale-free networks are prone to the spreading
  of rumors, just as they are to the spreading of infections. They are relevant
  to the spreading dynamics of chain emails, viral advertising and large-scale
  data dissemination algorithms on the Internet, the so-called gossip
  protocols.
\end{abstract}

\pacs{89.75.Hc,05.70.Jk,87.19.Xx, 89.75.Fb}

\maketitle

\section{INTRODUCTION}
Many real-life technological, social and biological systems 
have complex network-like structures, where vertices represent
entities comprising the systems and edges the presence of some form of
interaction or relationship between them.
Some important examples include the Internet \cite{f3,book1_internet}, 
the World  Wide Web \cite{web1}, and  
social interaction networks \cite{soc_nets4,
  soc_nets2,soc_nets3,soc_nets1}.
In recent years a significant amount of research has been devoted to 
empirical and theoretical  studies of such networks
\cite{review_albert, review_newman, review_mendes}.

A large body of this work has been focused on finding statistical properties,
such as degree distributions, average shortest paths, and clustering
coefficients that characterize the structure of complex networks.  Another
active area of research is on creating realistic models of such networks which
can help us better understand the underlying organization mechanisms behind
their formation and evolution.  A third area of research, which is still at
its early stages, investigates dynamical processes on complex networks with the
aim of understanding the impact of network topology on the dynamics
\cite{sis1_vesp,sir1_yamir,rumor1_zanette,rumor2_maziar, 
cascades_watts}.  

A prominent example
of such process is the spreading of rumors in social networks
\cite{book1_epidemic,DK1,DK2}.  Rumors bear an immediate resemblance with
epidemics and their spreading plays an important role in a variety of human
affairs, such as stock markets \cite{kimmel}, politics and warfare.
Rumor-like mechanisms also form the basis for the phenomena of viral marketing,
where companies exploit social networks of their customers on the Internet in
order to promote their products via the so-called `word-of-email' and
`word-of-web' \cite{viral1_domingos,viral2_domingos}.  Finally, rumor-mongering
forms the basis for an important class of communication protocols, called
gossip algorithms, which are used for large-scale information dissemination on
the Internet, and in peer-to-peer file sharing applications
\cite{gossip_demers,gossip2_review}.

Unlike epidemic spreading which has been extensively studied, quantitative
models and investigations of rumor dynamics have been rather limited.
Furthermore, most existing models either do not take into account the topology
of the underlying social interaction networks along which rumors spread, or use
highly simplified models of the topology.  While such simple models may
adequately describe the spreading process in small-scale social networks, via
the word-of-mouth, they become highly inadequate when applied to the spreading
of rumors in large social interaction networks, in particular those which are
formed on the Web.  Such networks, which include email networks
\cite{email_newman}, and social networking sites \cite{soc_nets1}, typically
number in tens of thousands to millions of nodes and their topology can show
highly complex connectivity patterns \cite{email_newman,soc_nets1}.

In two previous  papers \cite{rumor1_maziar,rumor2_maziar}
we investigated the dynamics of a
classical model of rumor spreading, the so-called Daley-Kendall model
\cite{DK1}, on such networks. In this paper we describe a more general model of
rumor dynamics on networks, which unifies our previous model with the
Susceptible-Infected-Removed (SIR) model of epidemics.  We formulate the
dynamics of this model within the framework of Interacting Markov Chains (IMC)
\cite{imc1}, and use this framework to derive a set of equations that describe
on a mean-field level the dynamics of the model on complex Markovian networks.
Using analytical and numerical solutions of these equations we examine both the
steady-state and the time-dependent behavior of the model on 
Erd\H os-R\'eyni (ER)
random graphs, as a prototypical example of homogeneous networks, and strongly
heterogenous scale-free (SF) networks, both in the absence and presence of
assortative degree correlations.

Our results point out to several important differences in the dynamics of rumor
spreading on the above networks. In particular, we find the presence of a
critical value for the rumor spreading rate (a rumor threshold) below which
rumors cannot spread in random graphs, and its absence in scale-free networks
with unbounded fluctuations in node degree distribution.  We find that the
initial spreading rate of a rumor is much higher on scale-free networks as
compared to random graphs, and that this spreading rate is further increased
when assortative degree-degree correlations are introduced. The final fraction
of nodes which ever hears a rumor (we call this the final size of a rumor),
however, depends on the interplay between the model parameters and the
underlying topology.  Our findings provide a better quantitative understanding
of the complex interplay between the characteristics of a rumor, i.e. its
spreading and cessation rates, and the topology of the underlying interaction
network that supports its propagation.  They are relevant to a number of
rumor-like processes such as chain emails, viral advertising and large-scale
data dissemination in computer and communication networks via gossip protocols.

The rest of this paper is organized as follows. In Section II we
describe our general rumor model. In section III a formulation of the
model  within the framework of Interactive Markov Chains is given, and 
the corresponding mean-field equations are derived. In section IV
analytical results are presented for the case of homogeneous networks, 
characterized by a constant degree and no degree correlations. This is
followed by a numerical study of the model on ER random graphs and
both uncorrelated and assortatively 
correlated SF networks. We close this paper in
section V with conclusions.

\section{A GENERAL RUMOR MODEL ON NETWORKS}

The model we shall consider is defined in the following way.  A closed
population consisting of $N$ members is subdivided into three groups, those who
have not heard the rumor, those who have heard it and wish to spread it, and
those who have heard it but have ceased communicating it. We call these groups
ignorants, spreaders and stiflers, respectively. The rumor spreads by directed
contact of the spreaders with others in the population, which takes place along
the links of an undirected social interaction network $G=(V,E)$, where $V$ and
$E$ denote the vertices and the edges of the network, respectively. The outcome
of contact of a specified spreader with (a) an ignorant is that the ignorant
becomes an spreader at a rate $\lambda$, and (b) another spreader or a stifler
is that the initiating spreader becomes stifler at a rate  $\alpha$. The
spreading process starts with one (or more) element(s) becoming informed of a rumor
and terminates when no spreaders are left in the population.  In the
Daley-Kendall rumor model and its variants stifling is the only mechanism that
results in cessation of rumor spreading.  In reality, however, cessation can
occur also purely as a result of spreaders forgetting to tell the rumor, or their
disinclination to spread the rumor anymore. Following a suggestion in
\cite{book1_epidemic}, we take this important mechanism into account by
assuming that individuals may also cease spreading a rumor spontaneously (i.e.
without the need for any contact) at a rate $\delta$.

\section{INTERACTIVE  MARKOV CHAIN MEAN-FIELD EQUATIONS}
We can describe the dynamics of the above model on a network  within the framework of
the Interactive Markov Chains (IMC).  The IMC  was  originally
introduced in mathematical sociology, as a means for
modeling social processes involving many  interacting actors (or agents)
\cite{imc1}. More recently they have been applied to the dynamics of 
cascading failures in electrical power networks \cite{imc2}, and 
the spread of malicious software (malware) on the Internet \cite{imc3}.
% and the modeling of signal processing on networks in living cells
%\cite{imc4}. 
An IMC consists of $N$ interacting nodes, with each node having a
state that evolves in time according to an internal Markov chain.  Unlike
conventional Markov chains, however, the corresponding internal transition
probabilities depend not only on the current state of a node itself, but also
on the states of all the nodes to which it is connected.  The overall system
evolves according to a global Markov Chain whose state space dimension is the
product of states describing each node. When dealing with large networks, the
exponential growth in the state space renders direct numerical solution of the
IMCs extremely difficult, and one has to resort to either Monte Carlo
simulations or approximate solutions.  In the case of rumor model, each
internal Markov chain can be in one of the three states: ignorant, spreader or
stifler. For this case we derive below a set of coupled rate equations which
describe on a mean-field level the dynamics of the IMC. We note that a similar
mean-field approach may also be applicable to other dynamical processes on
networks which can be described within the IMC framework.

%For tractability we shall assume here that the rate at which 
%a spreader node communicates a rumor to its neighbors 
%the  small time interval $[t,t+\Delta t]$ is given by $\mu \Delta t$, where 
%$\lambda$ is the rumor spreading rate. 

Consider now a node $j$ 
which is in the ignorant state at time $t$. We denote with $p^j_{ii}$
the probability that this node stays in the ignorant state in 
the time interval $[t,t+\Delta t]$, and with
$p^j_{is}=1-p^j_{ii}$ the probability that it makes a transition to
the spreader state. It then follows that 
\be
p_{ii}^{j}= (1-\Delta t \lambda)^{g},
\ee
where  $g=g(t)$ denotes the number of neighbors of node $j$
which are in the spreader state at time $t$. In order 
to progress, we shall coarse-grain the micro-dynamics of the system by
classifying  nodes in our network according to their degree  
and taking statistical average of the above transition probability
over degree classes. 

Assuming that a node $j$ has $k$ links, $g$  can be considered
as an stochastic variable which has the following binomial distribution: 
\be
\Pi(g,t)
=\binom{k}{g}\theta(k,t)^{g}(1-\theta(k,t))^{k-g},
\ee
where $\theta(k,t)$ is the probability at time $t$ 
that an edge emanating from 
an ignorant  node with $k$ links points to a spreader node. This 
quantity can be written as 
\be
\theta(k,t)= \sum_{k'} P(k'|k)P(s_{k'}|i_k) \approx 
\sum_{k'} P(k'|k)\rho^s(k',t) .
\ee
In this equation $P(k'|k)$ is 
the degree-degree correlation function, $P(s_{k'}|i_k)$ the 
conditional probability that a node with $k'$ links is in the spreader
state, given that it is connected to an ignorant node with degree $k$,
and $\rho^s(k',t)$ is the density of spreader nodes  at time
$t$ which belong to connectivity class $k$.  
In the above equation the 
final approximation is obtained by ignoring  dynamic correlations
between the states of neighboring  nodes. 

The transition probability $\bar{p}_{ii}(k,t)$ averaged over all possible 
values of $g$ is then given by: 
\begin{eqnarray}
\bar{p}_{ii}(k,t) & = &
\sum_{g=0}^k \binom{k}{g}
(1-\lambda\Delta t )^g\theta(k,t)^{g}(1-\theta(k,t))^{k-g} \nonumber \\
& = &  
\left( 1-\lambda \Delta t \sum_{k'}P(k'|k)\rho^s(k',t)\right)^k.
\end{eqnarray}

In a similar fashion we can derive an expression for the
probability 
$\bar{p}_{ss}(k,t)$ that a spreader node which has 
$k$ links stays in this state  in the interval
$[t,t+\Delta t]$. In this case, however, we also  need
to compute the expected value of the number of stifler neighbors
of the node at time $t$. Following steps  similar to the previous paragraphs
we obtain  
\\
\begin{eqnarray}
\bar{p}_{ss}(k,t) &=&
\left( 1-\alpha \Delta t
\sum_{k'}P(k'|k)(\rho^s(k',t)+\rho^r(k',t))\right)^k \nonumber \\
&\times&  (1-\delta \Delta t).
\end{eqnarray}
The corresponding probability for a transition from the spreader
to the stifler state, $\bar{p}_{sr}(k,t)$ is given by $
\bar{p}_{sr}(k,t)=1-\bar{p}_{ss}(k,t)$.

The above transition probabilities can be used to set up a system of 
coupled stochastic equations for change in time of the population of 
ignorants, spreaders and stiflers within each connectivity class 
\cite{maziar_unpub}.
However, ignoring fluctuations around expected values we can also 
obtain a set of deterministic rate equations for the expected values 
of these quantities. In the limit $\Delta t \rightarrow 0$ these 
equations are given by
\be
\frac{\partial \rho^i(k,t)}{\partial t}=-k
\lambda \rho^i(k,t)\sum_{k'}\rho^s(k',t)P(k'|k) 
\ee
\begin{eqnarray}
\frac{\partial \rho^s(k,t)}{\partial t} &=&
k \lambda \rho^i(k,t) \sum_{k'}\rho^s(k',t)P(k'|k)
\nonumber \\
&-& 
k\alpha \rho^s(k,t) 
\sum_{k'}(\rho^s(k',t)+\rho^r(k',t))P(k'|k) \nonumber \\
&-& \delta \rho^s(k,t)
\end{eqnarray}
\begin{eqnarray}
\frac{\partial \rho^r(k,t)}{\partial t} &=&
k \alpha \rho^s(k,t) 
\sum_{k'}(\rho^s(k',t)+\rho^r(k',t))P(k'|k) + \nonumber \\
&+& \delta \rho^s(k,t).
\end{eqnarray}

In the above equations $\rho^i(k,t),\rho^s(k,t)$, and $\rho^r(k,t)$
are the densities at time $t$ of, respectively, ignorants, spreaders and
stiflers in class $k$. These quantities  satisfy the normalization 
condition $\rho^{i}(k,t)+\rho^s(k,t)+\rho^r(k,t)=1$.

For future reference we note here that  information on the underlying network 
is incorporated in the above equations solely via the degree-degree
correlation function. Thus in our analytical and numerical 
studies reported in the next section we do not need to generate any 
actual network. All that is required is either an analytical expression
for $P(k'|k)$ or a numerical representation of  this quantity.

\section{RESULTS AND DISCUSSIONS}
\subsection{Homogeneous networks}

In order to understand some  basic features of our rumor model we first 
consider the case of homogeneous networks, in which degree
fluctuations are very small and there are no degree correlations. 
In this case the rumor equations become: 
\begin{eqnarray}
\frac{d\rho^i(t)}{dt} & = &-\lambda \bar{k}\rho^i(t)\rho^s(t) \\
\frac {d\rho^s(t)}{dt} & = & \lambda \bar{k}\rho^i(t)\rho^s(t) 
-\alpha \bar{k} \rho^s(t)(\rho^s(t)+\rho^r(t)) \nonumber \\
& - & \delta \rho^s(t) 
\\
\frac{d\rho^r(t)}{dt} & =& \alpha\bar{k}\rho^s(t)(\rho^s(t)+
\rho^r(t)) +\delta \rho^s(t),
\end{eqnarray}
where $\bar{k}$ denotes the constant degree distribution of the network (or the
average value for networks in which the probability of finding a node with a
different connectivity decays exponentially fast).

The above system of equations can be integrated analytically using an
standard approach. In the infinite time limit, when $s(\infty)=0$, we
obtain the following transcendal equation 
for $R= r(\infty)$, the final fraction of nodes which ever hear the
rumor (we call this the final rumor size) 
\be
R=1-e^{ -\epsilon R}
\ee
where 
\be
\epsilon= 
\frac{(\alpha+\lambda)\bar{k}}{\delta+\alpha\bar{k}}. 
\ee
Eq. (12)  admits a non-zero solution  only if
$\epsilon>1$. For $\delta \neq 0$ this condition is fulfilled provided 
\be
\frac{\lambda}{\delta}\bar{k} >1 ,
\ee
which is precisely the same threshold condition as found in the SIR
model of epidemic spreading on homogeneous networks 
\cite{sir1_yamir, sir1_ml}. 
On the other hand, in the special case $\delta=0$ (i.e when the
forgetting mechanism is absent) $\epsilon=1+\lambda/\alpha>1$, and  so Eq. (14) always admits a
non-zero solution, in agreement with the result in
\cite{rumor1_maziar}.

The above result shows, however, that the presence of a forgetting mechanism
results in the appearance of a finite threshold in the rumor spreading rate
below which rumors cannot spread in homogeneous networks. Furthermore, the
value of the threshold is {\em independent} of $\alpha$ (i.e. the stifling
mechanism), and is the same as that for the SIR model of epidemics on such
networks.  This result can be understood by noting that in the above equations
the terms corresponding to the stifling mechanism are of second order in $\rho^s$,
while the terms corresponding to the forgetting mechanism are only of first
order in this quantity. Thus in the initial state of spreading process, where
$\rho^s\approx 0$ and $\rho^r\approx0$, the effect of stifling is negligible relative to that of
forgetting, and the dynamics of the model reduces to that of the SIR model.

\subsection{Random graphs and uncorrelated scale-free networks}
Next we numerically investigate the dynamics of rumor spreading  on
complex networks. We consider first {\it uncorrelated}
networks, for which the degree-degree correlations can be written as 
\be
P(k'|k)= q(k')=\frac{k'P(k')}{\langle k \rangle},
\ee
where $P(k')$ is the degree distribution and $\langle k \rangle $ is the 
average degree.  We consider here two classes of such networks. 
The first class is the Erd\H os-R\'enyi random  networks, which  for large $N$  
have a Poisson degree distribution: 
\be
P(k)=e^{-k}\frac{\langle k \rangle ^k}{k!}.
\ee
The above degree distribution peaks at an average value $\langle k \rangle$ 
and show small fluctuations around this value. The second class we consider 
are scale-free networks which are 
characterized by a highly right-skewed power law degree distribution:
 \be
P(k)=
\begin{cases}
Ak^{-\gamma}
& k_{\text{min}}\leq k
\\
0                                                           & \text{otherwise.}
\end{cases}
\ee
In the above equation $k_{\text{min}}$ 
is the minimum degree of the networks and $A$
is a normalization constant.
For $2\leq \gamma \leq 3$ the variance of the above degree
distribution becomes infinite in the limit of infinite system size
while the average degree distribution remains finite.  
We shall consider henceforth SF networks with $\gamma=3$. 

Our studies of uncorrelated networks were performed using the above 
forms of $P(k)$ to represent  ER and SF networks, respectively.
The size of the networks considered was $N=10^5$ $N=10^6$, 
and the average degree was fixed at $ \langle k \rangle
=7$. For each network considered we generated a sequence of 
$N$ random integers distributed according to its 
degree distribution. The 
coupled set of differential equation (6-8) were then solved
numerically using an standard 
finite difference scheme, and  numerical  convergence 
with respect to the step size was checked numerically.
In the following and throughout the paper 
all calculations reported  are  performed
by starting the rumor from a randomly chosen initial spreader, and
averaging the results over 300 runs with different initial spreaders.
The calculations reported below were performed for networks consisting
of $N=10^6$ nodes.

In our first set of calculations we set $\delta=1$ and investigate 
the dynamics as a function of the rumor spreading rate $\lambda$
and the stifling rate $\alpha$.  
First we focus on the impact of network topology 
on  the final size of a rumor, $R$, which for inhomogeneous networks 
is obtained from 
\be
R=\sum_k\rho^r(k,t_{\infty}),
\ee
where $t_\infty$ denotes a sufficiently long time at which 
the spreading process has reached  its steady state 
(i.e. no spreaders are left in the network). In Fig. 1 $R$ 
corresponding to the ER network is 
plotted as a function of $\lambda$, and for 
several different values of the stifling parameter $\alpha$.
It can be seen that in this network $R$  exhibits a critical 
threshold $\lambda_c$ below which a rumor cannot spread in the
network. Furthermore, just as in the case of homogeneous networks,
the value of the threshold does not depend on $\alpha$, and is 
at $\lambda_c=0.12507$. This value is in excellent agreement 
with the analytical results for the SIR model on an infinite size ER
network, which is given by  $\lambda_c = \langle k \rangle 
/\langle k^2 \rangle =0.125$ \cite{sir1_yamir}. We also verified
numerically that the behavior of  $R$ in the vicinity of the critical 
point can be described with the form 
\be
R \sim A(\lambda-\lambda_c)^\beta,
\ee
where $\beta=1$, and $A=A(\alpha)$ is a smooth and monotonically decreasing
function of $\alpha$. The results are  shown in Fig. 2 where $R$ is plotted as 
function of $\lambda$ for a range of values of $\alpha$, together 
with the corresponding fits. 

We further examined  the above numerical findings 
by analytically solving  Eqs. (6-8) in the vicinity of the critical 
rumor threshold, to first order in $\alpha$. 
Details of the calculations are given in the Appendix, and 
they  confirm the above  numerical results for $\lambda_c$, and 
the behavior of $R$ in the vicinity of the critical rumor threshold.

Next we turn to our results for the SF network. 
In Fig. 3 results for $R$  in this  network are shown. In
this case we also observe the presence of an $\alpha$-independent 
rumor threshold, 
albeit for much smaller spreading rates than for the ER network.
We have verified  numerically that in this case the threshold is
approached with zero slope, as can also be gleaned from Fig. 3. 
Since the value of the threshold is independent of $\alpha$,
we can use the well-known result
for the SIR model (the $\alpha=0$ case) to conclude that in the
limit of infinite system size the threshold seen in the 
SF network will approach zero. It is therefore not an
intrinsic property of rumor spreading on this network.

\begin{figure}
 \epsfig{file=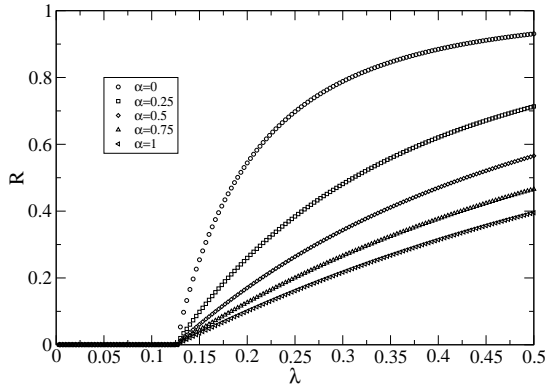,width=2.5in, angle=-90, clip=1} 
\caption{ The final size of the rumor, $R$ is shown 
as a function of the spreading rate $\lambda$ 
for the ER network of size $10^6$. The results are shown 
for several values of the stifling parameter $\alpha$.} 
\end{figure}

\begin{figure}
 \epsfig{file=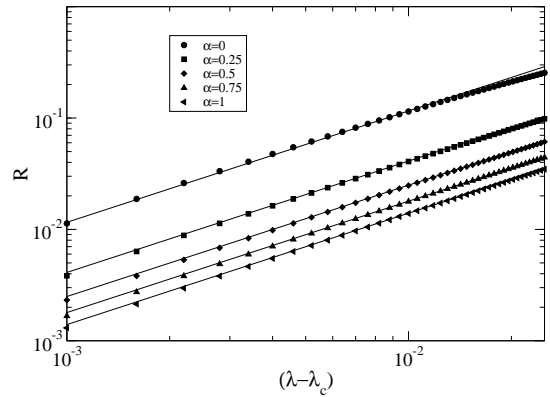,width=2.5in, angle=-90, clip=1}
 \caption{$R$ is plotted as a function of $\lambda-\lambda_c$ for the 
ER network of size $10^6$, using different values of $\alpha$. 
Solid lines show our numerical fits to the 
form $R\sim (\lambda-\lambda_c)^\beta$, with $\beta=1$.}
\end{figure}

\begin{figure} 
\epsfig{file=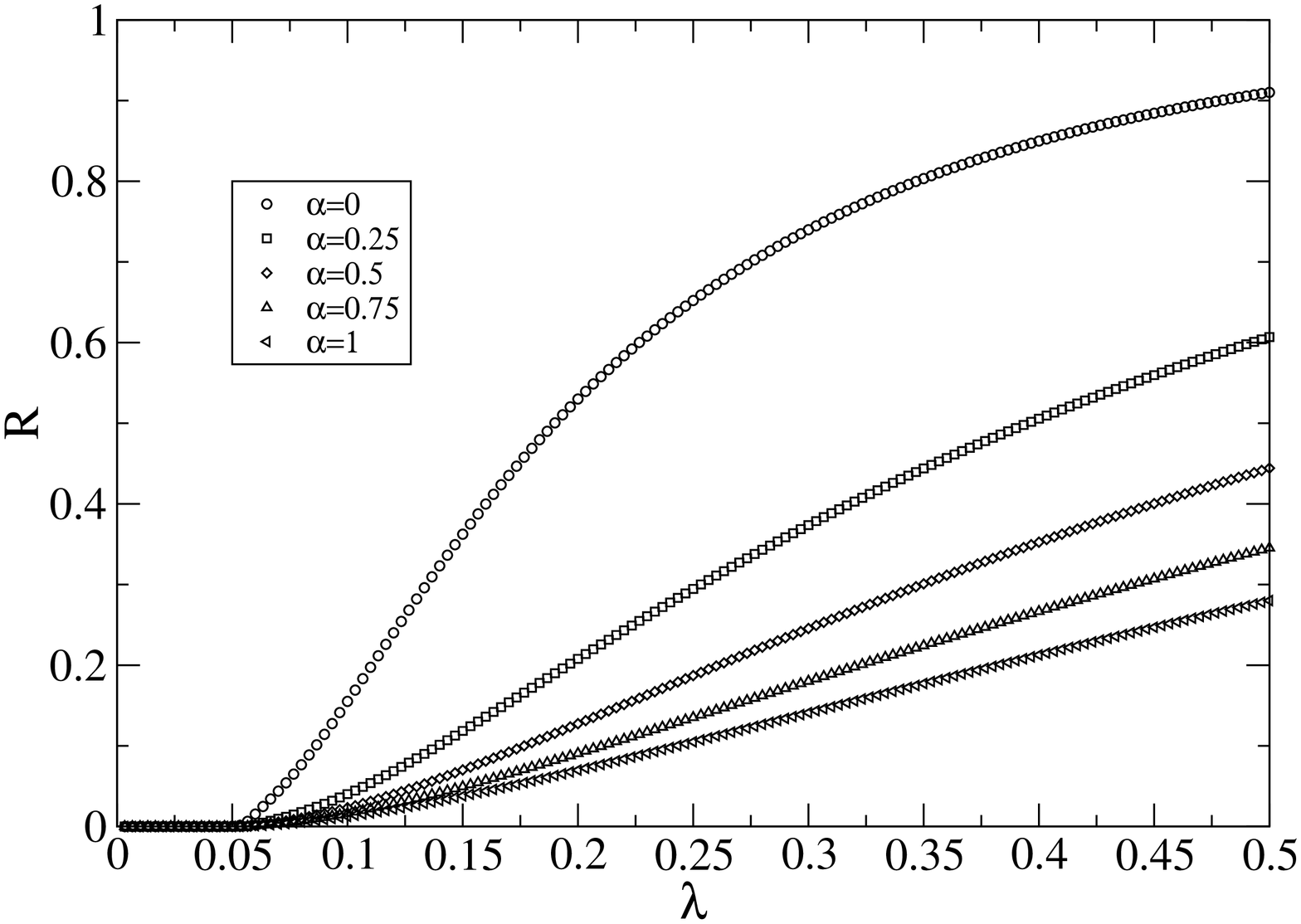,width=2.5in, angle=-90, clip=1}
 \caption{The final size of the rumor, $R$ is shown 
as a function of the spreading rate $\lambda$ 
for the SF network of size $10^6$. The results are shown 
for several values of the stifling parameter $\alpha$.}
\end{figure}

\begin{figure}
 \epsfig{file=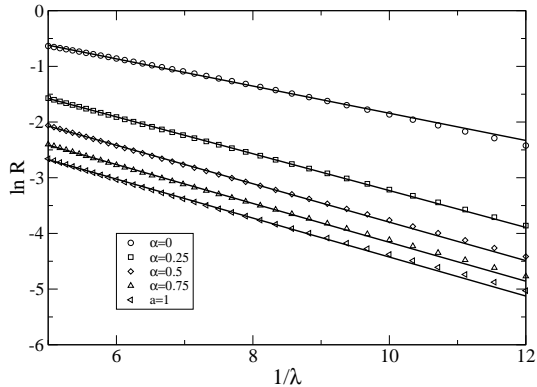,width=2.5in, angle=-90, clip=1}
 \caption{$R$ (in log scale) in the SF network of size $10^6$
   is plotted as a function of $1/\lambda$ and several values of $\alpha$. Solid lines are
   our numerical fits to the stretched exponential form $R= B(\alpha)\exp(-C(\alpha)/\lambda)$.
   The network size is $N=10^6$}
\end{figure}

In order to further analyze the behavior of $R$ in SF networks, we have
numerically fitted our results to the stretched exponential form, \be R \sim
\exp(-C/\lambda), \ee with $C$ depending only weakly on $\alpha$.
This form was found to describe the epidemic
prevalence in both the SIS and the SIR model of epidemics
\cite{sis1_vesp,sir1_yamir}.  The results are displayed in Fig. 4, and they
clearly indicate that the stretched exponential form also nicely describes the
behavior of $R$ in our rumor model. This result provides further support for
our conjecture that the general rumor model does not exhibit any 
threshold behavior
on SF networks (at least in the limit of infinite systems size).

In addition to investigating the impact of network topology on the steady-state
properties of the model, it is of great interest to understand how the
time-dependent behavior of the model is affected by topology.  In Figs. 5 and 6
we display, as representative examples, time evolution of, respectively, the
total fractions of stiflers and spreaders, in both networks for $\lambda=1$ and two
sets of values of the cessation parameters: $\{\delta=1,\alpha=0 \}$, and $\{\delta=0, \alpha=1\}$. The
first set of parameters corresponds to a spreading process in which cessation
results purely from spontaneous forgetting of a rumor by spreaders, or their
disinclination to spread the rumor any further.  The second set corresponds to
a scenario where individuals keep spreading the rumor until they become
stiflers due to their contacts with other spreaders or stiflers in the network.
As can be seen in Fig. 5, in the first scenario the initial spreading rate of
a rumor on the SF network is much faster than on the ER network. In fact, we
find that the time required for the rumor to reach $50\%$ of nodes in the ER
random graph is nearly twice as long as the corresponding time on the
SF networks.
This large difference in the spreading rate is due to the presence of highly
connected nodes (hubs) in the SF network, whose presence greatly speeds up the
spreading of a rumor. We note that in this scenario not only a rumor spreads
initially faster on SF networks, but it also reaches a higher fraction of nodes
at the end of the spreading process.

It can be seen from Figs, 5 and 6 that in the second spreading scenario 
(i.e. when stifling is the only mechanism for cessation) the 
initial spreading rate on the SF network is, once again, higher than
on the ER network. However, unlike the previous situation, the ultimate
size of the rumor is higher on the ER network. 
This behavior is due to the conflicting 
roles that hubs play when the stifling mechanism is switched
on. Initially the presence of hubs speeds up the spreading but once
they turn into stiflers they also effectively impede further spreading of the 
rumor.

\begin{figure}
 \epsfig{file=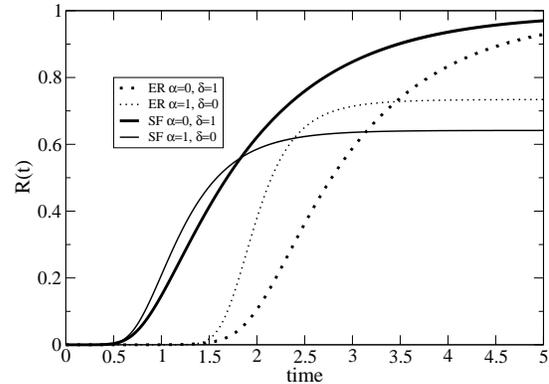,width=2.5in, angle=-90, clip=1} \\
 \caption{Time evolution of the density  of stiflers is shown on
   the ER (dashed lines) and the SF network (solid lines) when the dynamics
   starts with a single spreader node. Results are shown for two sets of values
   of the cessation parameters $\{\alpha=0,\delta=1$\} and $\{\alpha=1,\delta=0\}$. The network sizes
   are $N=10^6$.}
\end{figure}

\begin{figure}
 \epsfig{file=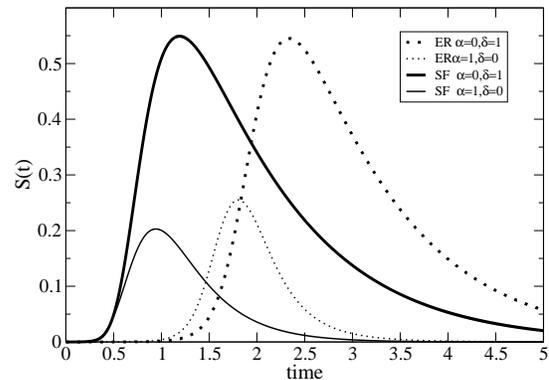,width=2.5in, angle=-90, clip=1} \\
 \caption{Time evolution of the density  of spreaders is shown  for
 the same networks, model parameters and initial conditions as in
 Fig. 5}
\end{figure}

\subsection{Assortatively correlated scale-free networks}

Recent studies have revealed that social networks display 
assortative degree  correlations,  
implying that highly connected vertices preferably connect to
vertices which are also highly connected 
\cite{review_newman}.  In order to study the impact of such
correlations on the dynamics of our model, we make 
use of the following ``local'' 
ansatz for the degree-degree  correlation function 
\be
P(k'|k) = (1-\beta)q(k')+\beta \delta_{kk'}; \; \; \; \; \; \; \;
(0\leq\beta<1).
\ee
This form has been used previously in recent studies of the SIR 
dynamics on correlated scale-free networks \cite{vazquez1, sir2_yamir},
and  allows us  to study in a controlled way the impact of 
degree correlations on the spreading of rumor.

Using the above degree-degree correlation function we numerically solved Eqs.
(6-8) for a SF network characterized by $\gamma=3$ and 
$<k>=7$. The network size was
fixed at $N=100,000$, and we used two values for the correlation parameter:
$\beta=0.2$ and $\beta=0.4$. Fig. 7 displays $R$ as a function of $\lambda$, and for
$\alpha=0.5,0.75,1$ (the value of $\delta$ was fixed at $1$).

It can be seen that below $\lambda\approx 0.5$ a rumor will reach a
somewhat smaller fraction of nodes on the correlated networks than on the
uncorrelated ones.  However for larger values of $\lambda$ this behavior reverses,
and the final size of the rumor in assortatively correlated networks shows a
higher value than in the uncorrelated network. We thus conclude that the
qualitative impact of degree correlations on the final size of a rumor depends
very much on the rumor spreading rate.  We also investigated the effect of
assortative correlations on the dynamics (temporal behavior) of rumor spreading
and found that such correlations slightly increase the initial rate of
spreading.

\begin{figure}
 \epsfig{file=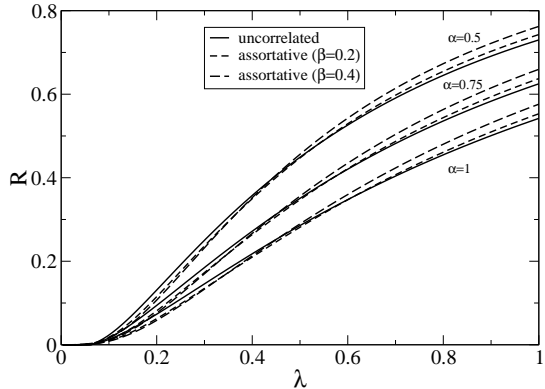,width=2.5in, angle=-90, clip=1} \\
\caption{The final size of the rumor  is plotted as a function of 
  $\lambda$ and for several values of $\alpha$ in the SF network of size $10^5$. Results
  are shown in the absence (solid lines) of assortative degree-degree
  correlations and in the presence of such correlations. The correlation
  strengths used are $\beta=0.2$ (short dashed lines) and $\beta=0.4$ (long dashed
  lines).}
\end{figure}

\section{CONCLUSIONS}

In this paper we introduced a general model of rumor spreading on complex 
networks. Unlike previous rumor models, our 
model incorporates two distinct mechanisms that cause 
cessation of a rumor, stifling and forgetting.
We used an Interactive Markov Chain formulation of the  model 
to derive deterministic mean-field equations
for the dynamics of the model on Markovian complex networks.
Using these equations we investigated analytically and numerically 
the behavior of the model on 
Erd\H os-R\'enyi random graphs and 
scale-free networks with exponent $\gamma=3$. 

Our results show the presence of a critical threshold in the rumor spreading
rate below which a rumor cannot spread in ER networks.  The value of this
threshold was found to be independent of the stifling mechanism, and to be the
same as the critical infection rate of the SIR epidemic model.  Such a
threshold is also present in the finite-size SF networks we studied, albeit at
a much smaller value. However in SF networks 
this threshold is reached with a zero slope and its
value becomes vanishingly small in the limit of infinite network size.  We also
found the initial rate of spreading of a rumor to be much higher on scale-free
networks than on ER random graphs. An effect which is caused by the presence of
hubs in these networks, which efficiently disseminate a rumor once they become
informed.  Our results show that SF networks are prone to the spreading of
rumors, just as they are to the spreading of infections.

Finally we used a local ansatz for the degree-degree correlation
function in order to numerically investigate the impact of assortative degree
correlations on the dynamics of our model on SF networks.  These correlations
were found to increase slightly the initial rate of spreading in SF networks.
However, their impact on the final fraction of nodes which hear a rumor depends
very much on the rate of rumor spreading.

The basic assumption underlying the mean-field equations derived in this paper
is that all vertices within a given degree class can be considered
statistically equivalent. Therefore our results are not directly applicable to
structured networks in which a distance or time ordering can be introduced, or
there is a high level of clustering.  We are currently working on more
elaborate approximations of our model which could take into account such
properties, and in particular the presence of clustering, which is known to be
an important feature of social networks.

Furthermore, in the present work we assumed the underlying network to be
static, i.e. a time-independent network topology.  In reality,
however, many social
and communication networks are highly dynamic. An example of such
time-dependent social networks is Internet chatrooms, where individuals
continuously make new social contacts and break old ones. Modeling the
dynamics of rumor spreading on such dynamic networks is highly challenging, in
particular when the time scale at which network topology changes becomes
comparable with the time scale of the process dynamics. We also aim to tackle
this highly interesting problem in future work.

\begin{acknowledgments}
M.~N. acknowledges the Abdus Salam International Centre for Theoretical 
Physics (ICTP) for a visiting fellowship 
during which some amendments  to this work were made. 
Y.~ M. is supported by MEC through the Ram\'{o}n y Cajal Program. This
work was supported by BT and the Spanish DGICYT Projects
FIS2004-05073-C04-01. We thank Keith Briggs for reading the manuscript.
\end{acknowledgments}

\appendix

\section{Critical rumor threshold in ER random graphs}
Eq. (6) can be integrated exactly to yield: 
\be
\rho^i(k,t)=\rho^i(k,0) e^{-\lambda k\phi(t)},
\ee
where $\rho^i(k,0)$ is the initial density of  ignorant nodes 
with connectivity $k$, and we have introduced the auxiliary function 
\be
\phi(t)=\sum_k q(k) \int_0^t
\rho^s(k,t')dt'\equiv \int_0^t \avg{\rho^s(k,t')}dt'.
\ee
In the above equation and hereafter we  use the shorthand notation
\be
\avg{O(k)}=\sum_k q(k) O(k)
\ee
with 
\be
q(k)=\frac{kP(k)}{\langle k \rangle}.
\ee
In order to obtain an expression for the final size of the rumor, $R$,
it is more convenient to work with $\phi$. Assuming an homogeneous
initial distribution of ignorants, $\rho^i(k,0)=\rho^i_0$,
we can obtain a differential equation for this quantity by multiplying 
Eq. (7) with $q(k)$ and summing over $k$. This yields after some
elementary manipulations:
\begin{eqnarray}
\frac{d\phi}{dt} &= & 1-\avg{e^{-\lambda k\phi}}) - 
\delta \phi \nonumber \\
& - & \alpha \int_0^t
\left[1-\avg{e^{-\lambda k\phi(t')}}\right] \avg{k\rho^s(k,t')}dt',
\end{eqnarray}
where, without loss of generality, we have also put $\rho^i_0\approx
1$.

In the limit $t\rightarrow \infty$  we have 
$\frac{d\phi}{dt}=0$, and Eq. (A5) becomes:
\begin{eqnarray}
0 & = & 1-\avg{e^{-\lambda k\phi_\infty}} 
- \delta\phi_\infty     \nonumber \\
&-&\alpha\int_0^\infty
\left[1-\avg{e^{-\lambda k\phi(t')}}\right] \avg{k\rho^s(k,t')}dt',
\nonumber \\
\label{phiinf}
\end{eqnarray}
where $\phi_\infty=\lim_{t\rightarrow \infty} \phi(t)$. 

For $\alpha=0$ Eq. (A5)  can be solved explicitly to obtain
$\Phi_{\infty}$ \cite{sir1_yamir}. 
For $\alpha\neq0$ we solve  (A5)  to leading order in 
$\alpha$. Integrating Eq. (7) to zero order in $\alpha$ we obtain
\be
\rho^s(k,t)=1-e^{-\lambda k\phi}-\delta \int_0^t e^{\delta(t-t')}
\left[1-e^{-\lambda k\phi(t')}\right]dt'+O(\alpha).
\ee
Close to the critical threshold both $\phi(t)$ and 
$\phi_\infty$ are small. Writing
$\phi(t)=\phi_\infty f(t)$, where $f(t)$ is a finite function, and
working to leading order in $\phi_\infty$, we obtain 
\be
\rho^s(k,t)\simeq -\delta \lambda k \phi_\infty \int_0^t
e^{\delta(t-t')}f(t') dt'+O(\phi_\infty^2)+O(\alpha)
\ee
Inserting this in Eq. (\ref{phiinf})  and expanding the
exponential to the relevant order in $\phi_\infty$ we find 
\begin{eqnarray}
0 & =& \phi_\infty\left[\lambda\avg{k}-\delta-
\lambda^2\avg{k^2}(1/2+\alpha\avg{k}I)\phi_\infty\right]
\nonumber \\
&+& O(\alpha^2)+O(\phi_\infty^3) 
\end{eqnarray}
where $I$ is a finite and positive-defined  integral. 
The non-trivial solution of this 
equation is  given by: 
\be
\phi_\infty=\frac{\lambda\avg{k}-\delta}
{\lambda^2\avg{k^2}(\frac{1}{2}+\alpha I\avg{k})}.
\ee
Noting that  $\avg{k}=\langle k^2 \rangle/\langle k \rangle$
and $\avg{k^2}=\langle k^3 \rangle /\langle k \rangle $
we obtain: 
\be
\phi_\infty=\frac{ 2\langle k \rangle
( \frac{\langle k^2 \rangle}{\langle k \rangle}\lambda -\delta) }  
{\lambda^2 \langle k^3 \rangle (1+ 2\alpha I
\frac{ \langle k^2 \rangle}{\langle k \rangle})}.
\ee
This yields a positive value for $\phi_{\infty}$ provided that 
\be
\frac{\lambda}{\delta} \geq \frac{\langle k \rangle}{\langle k^2 \rangle}.
\ee
Thus to leading order in $\alpha$ the critical 
rumor threshold is independent of this quantity, and is the same as 
for the SIR model.  In particular, for $\delta=1$ the critical rumor
spreading threshold is given by  
$\lambda_c= \langle k \rangle/\langle k^2 \rangle$, and 
Eq. (A12) simplifies to:

\be
\phi_\infty=\frac{ 2\langle k \rangle (\lambda -\lambda_c)}  
{\lambda^2 \langle k^3 \rangle(\lambda_c+2\alpha I)}.
\ee

Finally,  $R$ is given by 
\be
R=\sum_kP(k)(1-e^{-\lambda k\phi_\infty}),
\ee
and expanding the exponential in Eq. (A6) we obtain 
\begin{eqnarray}
R & \approx & \sum_k P(k) \lambda k \phi_\infty = 
\frac{ 2\langle k \rangle^2 (\lambda -\lambda_c)}  
{\lambda \langle k^3 \rangle(\lambda_c+2\alpha I)},
\end{eqnarray}
which shows that $R \sim (\lambda-\lambda_c)$ in the 
vicinity of the rumor threshold.

\end{document}